# Temperature dependent elastic constants and ultimate strength of graphene and graphyne

Tianjiao Shao[a, b], Bin Wen[a, b, 1], Roderick Melnik[c,d], Shan Yao[b], Yoshiyuki Kawazoe[e], Yongjun Tian[a]

[a]State Key Laboratory of Metastable Materials Science and Technology, Yanshan University, Qinhuangdao 066004, China

[b]School of Materials Science and Engineering, Dalian University of Technology, Dalian 116023, China

[c]M$^2$NeT Lab, Wilfrid Laurier University, Waterloo, 75 University Ave. West, Ontario, Canada N2L 3C5

[d]Ikerbasque, Basque Foundation for Science and BCAM, Bilbao 48011, Spain

[e]Institute for Materials Research, Tohoku University, 2-1-1 Katahira, Aoba-ku, Sendai 980-8577, Japan

**(February 03, 2012)**

**Abstract** Based on the first principles calculation combined with quasi-harmonic approximation, in this work we focus on the analysis of temperature dependent lattice geometries, thermal expansion coefficients, elastic constants and ultimate strength of graphene and graphyne. For the linear thermal expansion coefficient, both graphene and graphyne show a negative region in the low temperature regime. This coefficient increases up to be positive at high temperatures. Graphene has superior mechanical properties, with Young modulus $E_{11}$=371.0 N/m, $E_{22}$=378.2 N/m and ultimate tensile strength of 119.2 GPa at room temperature. Based on our analysis, it is found that graphene's mechanical properties have strong resistance against temperature increase up to 1200 K. Graphyne also shows good mechanical properties, with Young modulus $E_{11}$=224.7 N/m, $E_{22}$=223.9 N/m and ultimate tensile strength of 81.2 GPa at room temperature, but graphyne's mechanical properties have a weaker resistance with respect to the increase of temperature than that of graphene.

**PACS numbers:** 62.20.de, 65.40.-b

---

[1] Authors to whom any correspondence should be addressed
E-mail address: wenbin@ysu.edu.cn (Bin Wen), Tel: 086-335-8568761

# I. INTRODUCTION

New emerged two dimensional nano-films of graphene [1-4], graphyne and graphdiyne [5-9] have attracted many experimental and theoretical researchers, due to exceptional optical, electrical and mechanical properties of these materials. Since these two-dimensional materials are important applications for future in nano-electromechanical system (NEMS) and semiconductors, it is essential to clarify graphene and graphyne's mechanical properties and their resistance against temperature.

Recently, several works have been performed to calculate and measure the mechanical properties for single layer graphene. Already in 2001, Kudin et al predicted some mechanical properties of graphene by employing density functional theory (DFT) [13], in particular it was found that the elastic moduli along both armchair [1, 1] and zigzag [1, 0] directions are 345 N/m (1150.6 GPa by supposing the thickness of 0.335 nm). In 2007, Liu et al determined the phonon-induced instability by using first-principles method; they reported the Young modulus of 1050 GPa, ultimate strength of 110 GPa along the [1, 0] direction, and 121 GPa along [1, 1] direction, respectively [14]. In the same year, Khare et al investigated theoretically the deformation of graphene with defects through combining several approaches including the first principle method, molecular dynamics and a linear elastic continuum description [15]. In 2008, Lee et al measured graphene firstly in the experiment and reported the Young modulus of 340 N/m (1020 GPa) and ultimate strength of 130 GPa at room temperature [3]. In 2009, Wei et al evaluated the high order elastic constant of graphene up to the fifth order elastic constants by using DFT. In particular, they reported $C_{11}$=358.1 GPa and $C_{12}$=60.4 GPa. Their work added important insight into the nonlinear elastic behavior of graphene [17]. In 2010, Shen et al calculated the isotropic monolayer graphene's temperature dependent elastic constants (TDEC) at 300 K, 500 K and 700 K by employing classical molecular dynamics (MD) methodology, and both temperature and sample size effects on graphene were presented [19]. In the same year, Zhao et al predicted the graphene's temperature dependent ultimate strength (TDUS) from 0 K to 2500 K by using both classical MD and quantized fracture mechanics methods [20]. Although, mechanical properties of graphene have been measured and calculated, divergence in the actual values still exists, as shown in Table 1. In addition, temperature effects on the graphyne's mechanical properties have never been investigated to date, while the temperature influence on graphene's mechanical properties has only been evaluated by the classical MD where the quantum effect on nano-membranes was neglected.

The first principles analysis combined with the quasi harmonic approximation (QHA) method offer an accurate description for temperature dependent lattice dynamics for solids below Debye temperature. Very recently, a new methodology has been developed by our group for implementing lattice geometry optimization and calculating temperature dependent elastic constants (TDEC) within a given temperature range [21]. This work extends this methodology to address the problem of temperature dependent lattice geometry optimization and to determine mechanical properties of two dimensional lattices. Here, graphene and graphyne have been investigated under various simulation temperature ranges from 0 K to 1200 K. Based on this extension, in this work, temperature dependent lattice geometries, in-plane linear thermal expansion coefficients, temperature dependent elastic constants, Young moduli and ultimate strength of graphene and graphyne are analyzed and discussed in detail.

## II. METHODOLOGY

### A. Two dimensional lattice geometry optimization at given temperature

In our previous work, we have developed a new methodology for determining lattice geometries and temperature dependent elastic constants at high temperature for arbitrary symmetry three dimensional crystals [21]. The computation can be implemented by employing the first principles combined with the QHA. We further extend this theory and methodology to predict the lattice geometries and TDEC for two dimensional lattices.

Based on the crystallographic theory [22], the Gibbs energy is a function of the lattice lengths, lattice angles and all free crystallographic coordinates of the atoms in non-fixed Wyckoff positions and so forth. In this work, we suppose that free crystallographic coordinates of the atoms in non-fixed Wyckoff positions are fixed, then, configuration tensor X is only a function of lattice lengths $a$, $b$; lattice angle $\alpha$ of the two dimensional lattice. According to thermodynamics arguments [23, 24], non-equilibrium Gibbs energy for a specific phase can be written as

$$G[X(a,b,\alpha); p,T] = E[X(a,b,\alpha)] + pV[X(a,b,\alpha)] + A_{vib}[X(a,b,\alpha); T]. \qquad (1)$$

According to Eq. (1), the Gibbs energy is given as an energy surface in terms of a multi-variable function of $a$, $b$ and $\alpha$. To obtain the Gibbs energy at equilibrium, full minimization of Gibbs energy is required. One big challenge in this full minimization lies exactly with the fact that the Gibbs function is a multi-variable function, including $a$, $b$ and $\alpha$. In our previous work [21], we have proposed a scheme to address this issue by deriving the Gibbs energy in one variable $\xi$, allowing us to determine the anisotropy thermal expansion of this crystal and to obtain the linear directional thermal expansion coefficient.

By applying linear algebra and tensor analysis arguments [25], deformed two dimensional lattice configuration tensor can be expressed as the product of 4-dimensional deformation tensor and the initial configuration tensor, so that the deformed configuration tensor is given by,

$$X = X_0 \begin{pmatrix} e_1 & e_2 \\ e_3 & e_4 \end{pmatrix} \begin{pmatrix} \xi+1 & 0 \\ 0 & \xi+1 \end{pmatrix}, \tag{2}$$

where $X_0$ is the initial configuration tensor, $\begin{pmatrix} e_1 & e_2 \\ e_3 & e_4 \end{pmatrix}$ is a normal deformation tensor and $\xi$ is the deformation strain.

According to Wang's work [26], if the applied deformation matrix is symmetric, then the deformed configuration tensor can be further rewritten as

$$X = X_0 \begin{pmatrix} e_1 & e_2 \\ e_2 & e_3 \end{pmatrix} \begin{pmatrix} \xi+1 & 0 \\ 0 & \xi+1 \end{pmatrix}. \tag{3}$$

Then, the Gibbs energy is given as a unique function of the deformation strain $\xi$, and the non-equilibrium Gibbs energy is expressed as follows

$$G[X(a,b,\alpha); p,T] = E[X(\xi)] + pV[X(\xi)] + A_{vib}[X(\xi);T], \tag{4}$$

where $E[X(\xi)]$ is the total energy of the specific deformed configuration, $A_{vib}[X(\xi);T]$ is the vibrational Helmholtz free energy, which can be calculated from phonon density of states by using the QHA.

$$A_{vib}[X(\xi);T] = \int_0^\infty [\frac{1}{2}\hbar\omega + kT\ln(1-e^{-\hbar\omega/k_B T})]g[X(\xi);\omega]d\omega, \tag{5}$$

where $\omega$ is the phonon frequency, $T$ is the temperature, $k$ and $\hbar$ are the Boltzmann constant and the reduced Planck constant, respectively.

The Gibbs energy at equilibrium, which is only related to the parameter of temperature $T$, can be evaluated as the minimum of the non-equilibrium Gibbs energy with respect to the deformation strain $\xi$,

$$G^*(\xi_{p,T}^0; p,T) = \min_\xi \{E[X(\xi)] + pV[X(\xi)] + A_{vib}[X(\xi);T]\}. \tag{6}$$

When the deformation mode is fixed, the deformation strain $\xi_{p,T}^0$ at equilibrium along this deformation mode can be obtained, which is a function of lattice parameters $a$, $b$ and $\alpha$. To find the lattice parameter $a$, $b$ and $\alpha$ for a *P1* symmetry two dimensional lattice, a system of three equations is required to be solved simultaneously,

$$\begin{cases} F_1(a,b,\alpha,\xi_{1,p,T}^0) = 0, \\ F_2(a,b,\alpha,\xi_{2,p,T}^0) = 0, \\ F_3(a,b,\alpha,\xi_{3,p,T}^0) = 0, \end{cases} \quad (7)$$

where three equilibrium deformation strains $\xi_{1,p,T}^0$, $\xi_{2,p,T}^0$ and $\xi_{3,p,T}^0$ from three independent deformation modes are obtained to evaluate the independent lattice parameters $a$, $b$ and $\alpha$ for the two dimensional lattice.

Graphene and graphyne calculated in this work both belong to the hexagonal symmetry. To obtain the lattice geometry at a given temperature, the deformation tensor $\begin{pmatrix} \xi_1 & 0 \\ 0 & \xi_1 \end{pmatrix}$ is chosen to implement the lattice geometry optimization at this temperature. In this case, the Eq. (6) can be cast in the following form to obtain the lattice geometry at temperature $T$

$$G^*(\xi_{1,p,T}^0; p, T) = \min_{\xi_1} \left\{ E[X(\xi_1)] + pV[X(\xi_1)] + A_{vib}[X(\xi_1);T] \right\}. \quad (8)$$

We solve the Eq. (8) to obtain the deformation strain $\xi_{1,p,T}^0$ at equilibrium at given temperature $T$. Then the lattice length for hexagonal graphene or graphyne can be obtained as

$$a_{p,T} = (1 + \xi_{1,p,T}^0)a_0, \quad (9)$$

where $a_0$ represents the optimized hexagonal two-dimensional lattice parameter $a$ at 0 K, which can be determined from the first-principles calculation; $a_{p,T}$ is the lattice parameter $a$ at equilibrium at given temperature $T$ and pressure $p$.

## B. Two-dimensional lattice's TDEC

Based on our developed methodology for calculating the TDEC for three dimensional solid [21], we have derived the Helmholtz free energy for evaluating the temperature dependent elastic constants

$$F[X(a,b,\alpha); p,T] = E[X(\xi)] + A_{vib}[X(\xi);T], \qquad (10)$$

where $E[X(\xi)]$ is the total energy for the specific deformed configuration, $A_{vib}[X(\xi);T]$ is the vibrational Helmholtz free energy, which can be calculated by first-principles approaches combined with QHA model.

Since all the stress components out of the two-dimensional plane equal to zero, for the two dimensional lattice the dimension of elastic constant matrix is reduced to 3 and is given by,

$$C = \begin{pmatrix} c_{11} & c_{12} & c_{16} \\ c_{12} & c_{22} & c_{26} \\ c_{16} & c_{26} & c_{66} \end{pmatrix}. \qquad (11)$$

On the basis of continuum elasticity theory [27, 28], the elastic constants are given by the second derivative of the Helmholtz free energy with respect to the deformation strain tensor. For evaluating the 6 independent elastic constants $c_{ij}^{p,T}$ (*i, j*=1, 2 or 6), a system of equations that correspond to the 6 independent deformation modes are required to solve.

$$\begin{cases} F_1(D_1^{p,T}, c_{11}^{p,T}, c_{12}^{p,T}, c_{22}^{p,T}, c_{16}^{p,T}, c_{26}^{p,T}, c_{66}^{p,T}) = 0 \\ F_2(D_2^{p,T}, c_{11}^{p,T}, c_{12}^{p,T}, c_{22}^{p,T}, c_{16}^{p,T}, c_{26}^{p,T}, c_{66}^{p,T}) = 0 \\ F_3(D_3^{p,T}, c_{11}^{p,T}, c_{12}^{p,T}, c_{22}^{p,T}, c_{16}^{p,T}, c_{26}^{p,T}, c_{66}^{p,T}) = 0 \\ F_4(D_4^{p,T}, c_{11}^{p,T}, c_{12}^{p,T}, c_{22}^{p,T}, c_{16}^{p,T}, c_{26}^{p,T}, c_{66}^{p,T}) = 0 \\ F_5(D_5^{p,T}, c_{11}^{p,T}, c_{12}^{p,T}, c_{22}^{p,T}, c_{16}^{p,T}, c_{26}^{p,T}, c_{66}^{p,T}) = 0 \\ F_6(D_6^{p,T}, c_{11}^{p,T}, c_{12}^{p,T}, c_{22}^{p,T}, c_{16}^{p,T}, c_{26}^{p,T}, c_{66}^{p,T}) = 0 \end{cases} \qquad (12)$$

where $D_i^{p,T}$ is the second order derivative of the Helmholtz free energy with respect to the deformation strain under the *i*-th deformation mode.

For the hexagonal graphene and graphyne lattices, due to the intrinsic symmetry, the elastic constant matrix is simplified to

$$C = \begin{pmatrix} c_{11} & c_{12} & 0 \\ c_{12} & c_{11} & 0 \\ 0 & 0 & (c_{11}-c_{12})/2 \end{pmatrix}. \tag{13}$$

Based on the Eq. (9) and Eq. (11), we choose the $\begin{pmatrix} \xi_1 & 0 \\ 0 & 0 \end{pmatrix}$ and $\begin{pmatrix} \xi_2 & 0 \\ 0 & \xi_2 \end{pmatrix}$ as the deformation modes to implement calculations the elastic constants for $c_{11}^{p,T}$ and $c_{12}^{p,T}$.

In the case of hexagonal graphene and graphyne, due to their anisotropic hexagonal symmetry, Young moduli along [10] direction and [01] are usually different. In particular, $E_{11}$ along [10] direction is given by,

$$E_{11} = C_{11} = \frac{1}{V_0(p,T)} \frac{\partial^2 F(p,T)}{\partial \xi} \tag{14}$$

where $\xi$ is the deformation strain along [1, 0] direction for the deformation mode $\begin{pmatrix} \xi_1 & 0 \\ 0 & 0 \end{pmatrix}$, while, Young modulus in [0, 1] direction is given by,

$$E_{22} = \frac{1}{V_0(p,T)} \frac{\partial^2 F(p,T)}{\partial \xi} \tag{15}$$

where $\xi$ the deformation strain along [0, 1] direction for the deformation mode $\begin{pmatrix} 0 & 0 \\ 0 & \xi_2 \end{pmatrix}$.

For the two deformation modes, the second order derivative of the Helmholtz free energy with respect to the deformation strain corresponds to the independent elastic constants or their linear combination. The corresponding relationship is given in Table 2.

## C. Two dimensional lattice's TDUS

In recent years, ultimate or ideal strengths have been calculated for various three dimensional crystals by using the first principle method [29, 30]. Most results have been obtained for three dimensional crystal's ideal strength at 0 K. To shed light on temperature's influence in graphene and graphyne and subsequent limitations on their tensile mechanical properties, we extend our methodology to evaluate the TDUS for these materials.

In order to evaluate the ultimate strength, we first need to find out the relationship for the lattice configuration under deformation strain versus the strain $\xi$. We obtain the equilibrium lattice configuration $X_0^{p,T}$ at temperature $T$ and stretch the unit cell in each step with a constant increase of $\Delta\xi$ (in our calculations $\Delta\xi$ =0.01 for the hexagonal graphene lattice and 0.02 for the hexagonal graphyne lattice). Then, we fix the length along the direction of tension, relax the unit cell to ensure that the other intrinsic stress perpendicular to the direction of tension is zero. Based on the above two steps, we obtain the configuration tensor $X(\xi)^{p,T}$ for the two-dimensional lattice under the tensile strain $\xi$ at given temperature $T$.

Recall that the Helmholtz free energy for the configuration lattice $X(\xi)^{p,T}$ is given by Eq (10). To calculate the stress versus the strain, we differentiate the Helmholtz free energy with respect to the strain [30]

$$\sigma_{tensile}(\xi, p, T) = \frac{1+\xi}{V(\xi)^{p,T}} \frac{dF(\xi, p, T)}{d\xi} \tag{16}$$

where $\sigma_{tensile}(\xi, p, T)$ is the tensile stress of the lattice under the tensile strain of $\xi$, $V(\xi)^{p,T}$ is the volume of the lattice under tensile strain of $\xi$, $F(\xi, p, T)$ is the Helmholtz free energy of the deformed lattice, $p$ is the pressure and $T$ is the temperature. Then, along a specific tensile deformation direction, the tensile stress $\sigma_{tensile}(\xi, p, T)$ with respect to the tensile strain $\xi$ at given pressure $p$ and given temperature $T$ can be calculated. Since ultimate strength is defined as the maximum stress that any material will withstand before destruction, by plotting the stress-strain curve, we determine the yielding point as the ultimate strength.

## D. Computational details

In this section, we provide details on the calculation of TDEC and TDUS for graphene and graphyne. Figure 1 shows the structure for the graphene and graphyne. Figure 1(a) shows graphene and its conventional graphene lattice containing 4 atoms, and circled with red solid line on this picture. The structure in Figure 1(b) is the graphyne super-cell structure and its conventional lattice containing 24 atoms which is circled with red solid line. To avoid the interaction between adjacent atomic sheets, the conventional lattice height Z for graphene and graphyne is set to be 15 Å and 10 Å, respectively.

The calculations are implemented in the Vienna ab-initio simulation package (VASP) developed by the Hafner Research Group [31]. We perform our first-principle calculation by using the plane-wave basis soft VASP projector augmented wave (PAW) method [32], within the local density approximation (LDA) [33, 34]. To determine the Helmholtz free energy of the configuration, we calculate the phonon spectra by using the displacement method [35] under the quasi-harmonic model [36, 37]. A plane-wave energy cutoff of 375 eV has been used and the Brillouin zone of the conventional graphene lattice, which has been sampled by $6 \times 4 \times 1$ **k**-point mesh for graphene conventional cell, including four atoms, $4 \times 2 \times 1$ **k**-point mesh for the graphyne conventional cell. Supercells of $2\times2$ times the graphene conventional cell including 16 atoms, $1\times1$ times the graphyne conventional cell including 24 atoms were used when computing the force constants. Phonon calculations were performed by using the PHONOPY code [36, 37].

# III. GRAPHENE'S MECHANICAL PROPERTIES

## A. Graphene's temperature dependent lattice geometry

In order to pave the way for the graphene's applications in graphene based devices where temperature changes in working surroundings, it is important to know the thermal expansion coefficient (TEC) for the graphene. During recent, several research works have been carried out to determine the TEC for the graphene by theoretical and experimental approaches [38-43]. In this work, we employ our methodology based on the first principle method combined with QHA to calculate the thermal expansion and lattice geometry characteristics for graphene at given temperature including quantum effect. The temperature dependent linear TEC for graphene that has been obtained is shown by the blue dot line in Figure 2(a). Below 430 K, this is the negative TEC region for graphene. Above the 430 K, the TEC for graphene is positive and increases slowly, reaching a small value of $2.53 \times 10^{-6} K^{-1}$ at 1200 K. These findings related to negative values of TEC at low temperatures and small positive values of TEC at high temperatures are in qualitative agreement with the previous measured and calculated data for free-standing graphene. In Figure 2 (a), the TEC measured by Bao et al are plotted by pink solid line [41], the values measured by Yoon et al are plotted by purple dash-dot line [43]; the values calculated by first principles by Mounet et al are shown by red dash line [38], those calculated by the non-equilibrium green function (NGF) approach are plotted by green dash-dot line [39], and are given for comparisons. The TECs of graphene measured in Bao et al [41] and Yoon et al's [43] by different experimental methods have common that the minimum value of TECs in both cases approaches $-1 \times 10^{-5} K^{-1}$. The negative minimum for the TEC of graphene shown by experiments is larger than our theoretical prediction and it might be contributed to the ripples in the real two dimensional films [41, 42]. This ripple in the graphene's thermal expansion process is clearly observed in experiments [41], as well as in AIMD simulations [42]. The reason for graphene's initial negative TEC region lies with its intrinsic two dimensional configuration property [41, 44].

Next, we plotted the graphene's lattice parameter $a$ by integrating the TEC versus graphene' lattice parameter $a$ from $a_0$ at 0 K as shown in Figure 2(b). The lattice parameter is $a = 2.4630$ Å at 0 K, then graphene shows a thermal contraction and achieves its minimum of lattice parameter of $a = 2.4626$ Å at 430 K, as shown by our first principle calculation. By increasing temperature further, graphene has a positive thermal expansion trend and its lattice parameter $a$ achieves the values of 2.4655 Å at 1200 K. Comparison is also made with the lattice parameter $a$ calculated by the *ab initio* molecular dynamic method (AIMD) from Pozzo et al's work as plotted by blue scatters [42], as well as with values calculated by the Monte carlo method from Zakharchenko et al's work as plotted by red solid circles [40]. The lattice parameter $a$ predicted in this work qualitatively agrees with these previous results.

## B. Graphene's TDEC

Graphene's Debye temperature is measured as being 1045 K and calculated to be 1287 K by Politano et al [45]. The QHA provides an effective and accurate tool for determining thermal characteristics below the Debye temperature. Therefore, in this work, we only report the graphene and graphyne's elastic constants and ultimate strengths versus temperature ranging from 0 K to 1200 K under the QHA. Experimental data and results obtained by other theoretical method are also taken as a comparison and they are shown in Figure 3(a). It is shown that $c_{11}$, which is obtained by deforming the hexagonal graphene lattice along the [1, 0] direction, is 373.0 N/m at 0 K (which corresponds to 1244.0 GPa by supposing the thickness of 0.335 nm), 371.10 N/m at 300 K (1237.6 GPa); while $c_{12}$= 40.33 N/m at 0 K (135.1 GPa), 37.95 N/m at 300 K (127.1 GPa). Elastic constants $c_{11}$ and $c_{12}$ in this work agree well with $c_{11}$=358.1 N/m; $c_{12}$=60.4 N/m at 0 K calculated by ab-initio method from Wei et al's work [17].

We obtain $E_{11}$=373.0 N/m (1223.1 GPa) at 0 K, 371.10 N/m (1237.6 GPa) at 300 K; $E_{22}$=380.3 N/m (1274.0 GPa) at 0 K, 378.2 N/m (1267.0 GPa) at 300 K, $E_{22}$ is slightly higher than Young modulus $E_{11}$. Moreover, the calculated Young modulus in this work is slightly larger than measured 340 N/m (1020 GPa) by Lee et al [3]. Firstly, this might be attributed to a perfect crystalline structure and periodical condition used in our first-principle calculations; secondly, the LDA has a tendency to overestimate the elastic modulus [35]. By using the MD approach, Shen et al calculated $c_{11}$ softening slope $\Delta c_{11}/T$ =0.0275 N/m; the $c_{12}$ softening slope $\Delta c_{12}/T$ =0.0425 N/m from 300 K to 700 K [13]; while in our work $\Delta c_{11}/T$ =0.0128 N/m, $\Delta c_{12}/T$ =0.0143 N/m in the corresponding temperature range. The relationship of this elastic constant versus temperature is in qualitative agreement with their work. By increasing temperature from 0 K to 1200 K, $c_{11}$ ($E_{11}$) decreases from 373.0 N/m to 359.7 N/m by 3.5%; $E_{22}$ decreases from 380.3 N/m to 363.1 N/m by 4.5%. It indicates that graphene can keep its superior mechanical properties even at high temperatures. Comparisons are also made with 345 N/m calculated by the ab-initio method from Liu et al's work [14], as well as with $c_{11}$=358.1 N/m, $c_{12}$=60.4 N/m calculated by the ab-initio method in Wei et al's work [17].

## C. Graphene's TDUS

Graphene's temperature dependent stress-strain relations at 0 K, 300 K, 800 K and 1200 K are plotted in Figure 3(c). The ultimate strength of 120.2 GPa in [1, 0] direction at 0 K predicted by our work is close to the ultimate strength of 121 GPa along the [1, 1] direction and 110 GPa along the [1, 0] direction calculated by Liu et al using the first-principle phonon calculation [14]. Beyond the deformation strain of 0.10, the stress depends linearly on the deformation strain, when the deformation strain surpasses the 0.10, a nonlinear mechanical behavior of the graphene was clearly shown. We further present the graphene's TDUS against temperature. When the temperature increases, the ultimate strength decreases with the increase of temperature. The ultimate strength is 120.2 GPa at 0 K, decreases to 114.4 GPa at 1200 K by 4.8 %. This download tendency in the elastic constant is caused by weaker interactions between the atoms induced by stronger vibrations due to increase of temperature. The ultimate strength of 119.2 GPa at 300 K is slightly lower than the 130 GPa measured by the Lee et al [3]. From 0 K to 1200 K, our calculated ultimate strength decreases with the increase of temperature monotonously, which is in qualitative agreement with the calculated results from Zhao et al's work by using the MD [20]. Our calculated results are qualitatively larger than those in Zhao et al's, because the ultimate strength decreases when the nano-film is under higher strain-rates. In our first principle calculation, the strain-rate is treated as static. This quasi-static strain-rate causes a larger ultimate strength compared to the results from simulations by the MD [20].

## IV. GRAPHYNE'S MECHANICAL PROPERTIES

### A. Graphyne's temperature dependent lattice geometry

Here, for the first time we present the temperature dependent lattice geometry and TEC for graphyne. In Figure 4 (a), the TEC for the graphyne is given, the graphyne is contracted with the increase of temperature and shows a negative TEC region between 0 K to 217 K, achieves its minimum negative value of $-0.36 \times 10^{-6} K^{-1}$ at 93 K. From 93 K, the TEC of graphyne shows a positive slope versus the temperature and begins to expand with the increase of temperature after 217 K; achieves the value of $0.61\times10^{-6} K^{-1}$ at 300 K and $5.37\times10^{-6} K^{-1}$ at 1200 K. The initial negative region for the TEC and then a positive thermal expansion with the increase of temperature is very similar to graphene's. This initial negative TEC region for graphyne is also induced by its intrinsic two dimensional configuration property [41, 44].

## B. Graphyne's TDEC

Next, we use our first principle calculation to predict the TDEC for hexagonal graphyne. Graphyne, which can be seen as the simplest material that belongs to the group of graphdiyne materials, also has superior mechanical properties as shown in our calculations. In Figure 4(b), $c_{11}$ ($E_{11}$) =239.0 N/m (which corresponds to 800.6 GPa by supposing the thickness of 0.335 nm), $c_{12}$=64.5 N/m (216.1 GPa), Young modulus $E_{22}$=239.0 N/m (800.7 GPa) at 0 K. The $c_{11}$ ($E_{11}$) of 239.0 N/m and Young modulus $E_{22}$ of 239.0 N/m are clearly smaller than their graphene's counterparts, e. g. $c_{11}$ ($E_{11}$) of 373.0 N/m and Young modulus of 380.3 N/m for grapheme at 0 K. From the optimized conventional cell at 0 K, we suppose their thickness being h=0.335 nm, graphyne's volume per carbon atom is given 11.52 $A^3$/ per carbon atom, while graphene's volume per carbon atom is given by 8.79 $A^3$/ per carbon atom. Relatively weaker atomic bond energy in graphyne, resulted from its more loose structure, might contribute to its weaker elastic properties. When temperature increases, the elastic constants of graphyne decrease substantially, e. g. at 1200 K, $c_{11}$ ($E_{11}$) =180.1 N/m ( 603.3 GPa ), $c_{12}$ = to 46.9 N/m (157.1 GPa), Young modulus $E_{22}$ =180.1 N/m (603.3 GPa). From 0 K to 1200 K, $c_{11}$ ($E_{11}$) softenes significantly by 24.8% and Young modulus $E_{22}$ decreases by 24.6%, respectively. These results suggest that graphyne's mechanical properties have weaker resistance with respect to the increase of temperature compared with graphene.

## C. Graphyne's TDUS

For hexagonal graphyne, we stretched the lattice along the x-axis as plotted in Figure 1(b); the stress-strain relationship at 0 K, 300 K, 800 K and 1200 K are shown in Figure 4(c). The ultimate strength for graphyne is clearly lower than for graphene at same given temperature. The same reason might be behind the lower Young modulus for graphyne due to weaker interactions between atoms originated from their loosen lattice configuration. Figure 4 (d) shows the TDUS for graphyne from 0 K to 1200 K. The tensile TDUS along the x-axis decreases monotonously with the temperature. By increasing temperature from 0 K to 1200 K, the ultimate strength for graphyne drops from 83.03 GPa to 72.5 GPa by 12.6%.

## V. SUMMARY AND CONCLUSIONS

This work has been motivated by the importance of two-dimensional nano-films' in a wide range of current and potential developments in NEMS and semiconductor technologies. Recently, in our previous work a new promising methodology for calculating temperature dependent lattice geometries and TDEC of three dimensional crystals has been proposed. Here, we have addressed in detail the problem of temperature dependent lattice geometry optimization, determining thermal extension coefficients, and calculating TDEC and TDUS for two dimensional graphene and graphyne by extending this methodology. Firstly, we found that the thermal expansion coefficient for both graphene and graphyne has a negative region at low temperatures and increases up to positive values at high temperatures. Graphene is shown to have exceptional mechanical properties, with Young modulus being $E_{11}$=371.0 N/m, $E_{22}$=378.2 N/m and ultimate tensile strength of 119.2 GPa at room temperature. These results agree well with recent experimental data of Young modulus of 340 N/m and ultimate strength of 130 GPa reported by Lee et al [3]. Graphene keeps its superior mechanical property even at elevated temperatures. We have reported here that the graphene's Young modulus of $E_{11}$ and $E_{22}$ only decrease by 3.5% and 4.5%, respectively; while its ultimate tensile strength drops by 4.8% from 0 K to 1200 K.

For graphyne, we have applied a similar methodology for temperature-dependent calculations. Graphyne's Young modulus is $E_{11}$=224.7 N/m, $E_{22}$=223.9 N/m and ultimate tensile strength is 81.2 GPa at room temperature. This indicates that graphyne, as the simplest material that belongs to the group of graphdiyne materials, has also strong mechanical properties, but inferior to graphene. Nevertheless, graphyne's mechanical properties soften significantly with the increase of temperature. Young modulus $E_{11}$ and $E_{22}$ soften by as much as 24.8% and 24.6%, respectively; while ultimate tensile strength drops from 83.03 GPa to 72.5 GPa by 12.6% from 0 K to 1200 K.

Finally, this work demonstrates a methodology for calculating two dimensional crystal's TDEC and TDUS and paves the way for an accurate prediction of two dimensional nano-film's lattice temperature dependent thermal and mechanical properties.

## ACKNOWLEDGMENTS

This work was supported by the National Natural Science Foundation of China (Grant No.'s 51121061, 51131002) and the Program for New Century Excellent Talents in Universities of China (NCET-07-0139). R. M. acknowledges the support from the NSERC and CRC programs, Canada. The authors also acknowledge the staff of the Center for Computational Materials Science, Institute for Materials Research, Tohoku University, for computer use.

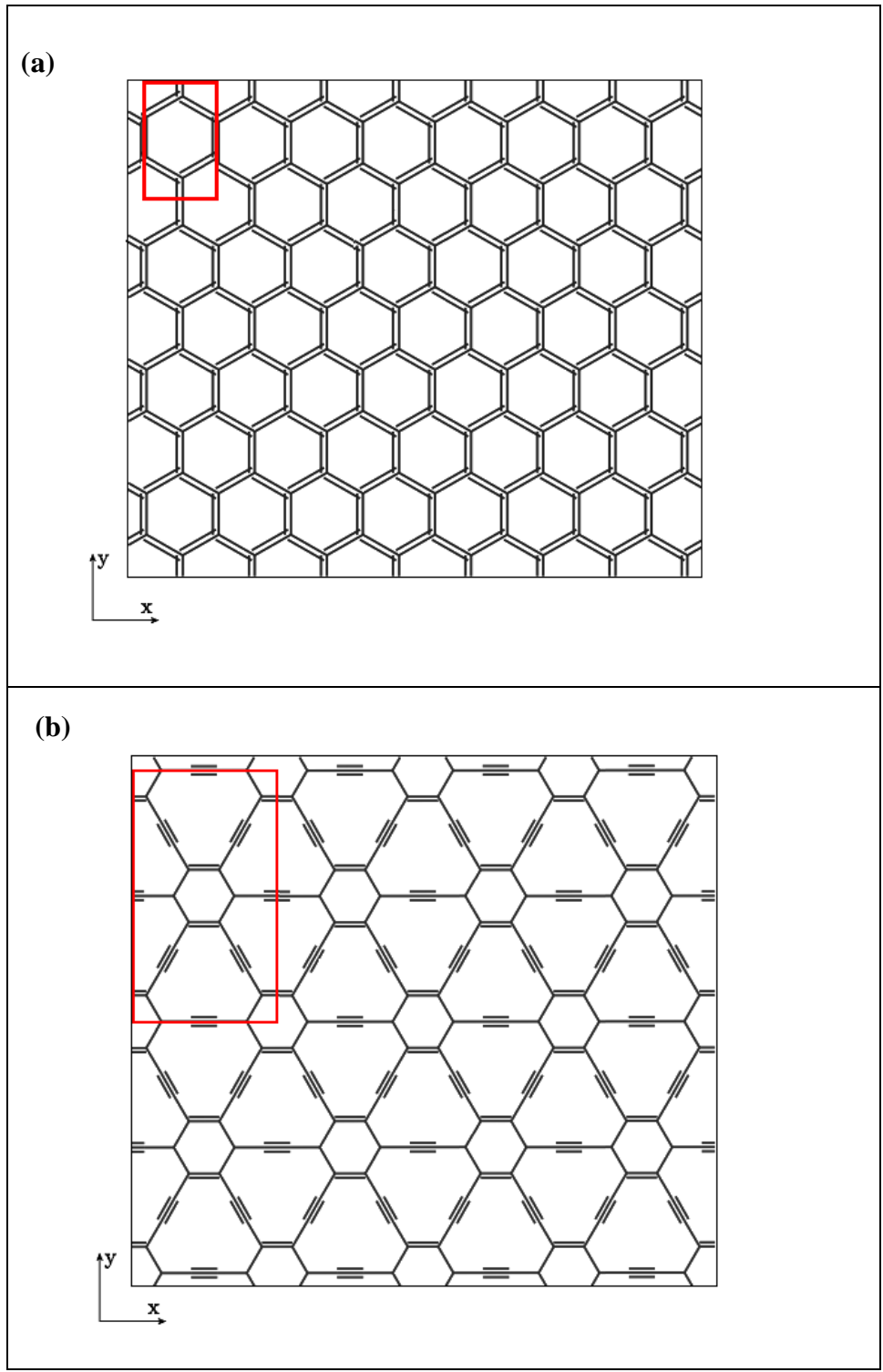

**Figure 1**. (a) Graphene and its four-atom conventional cell for calculation (circled with red line); (b) Graphyne and its 24-atom conventional cell for calculation (circled with red line).

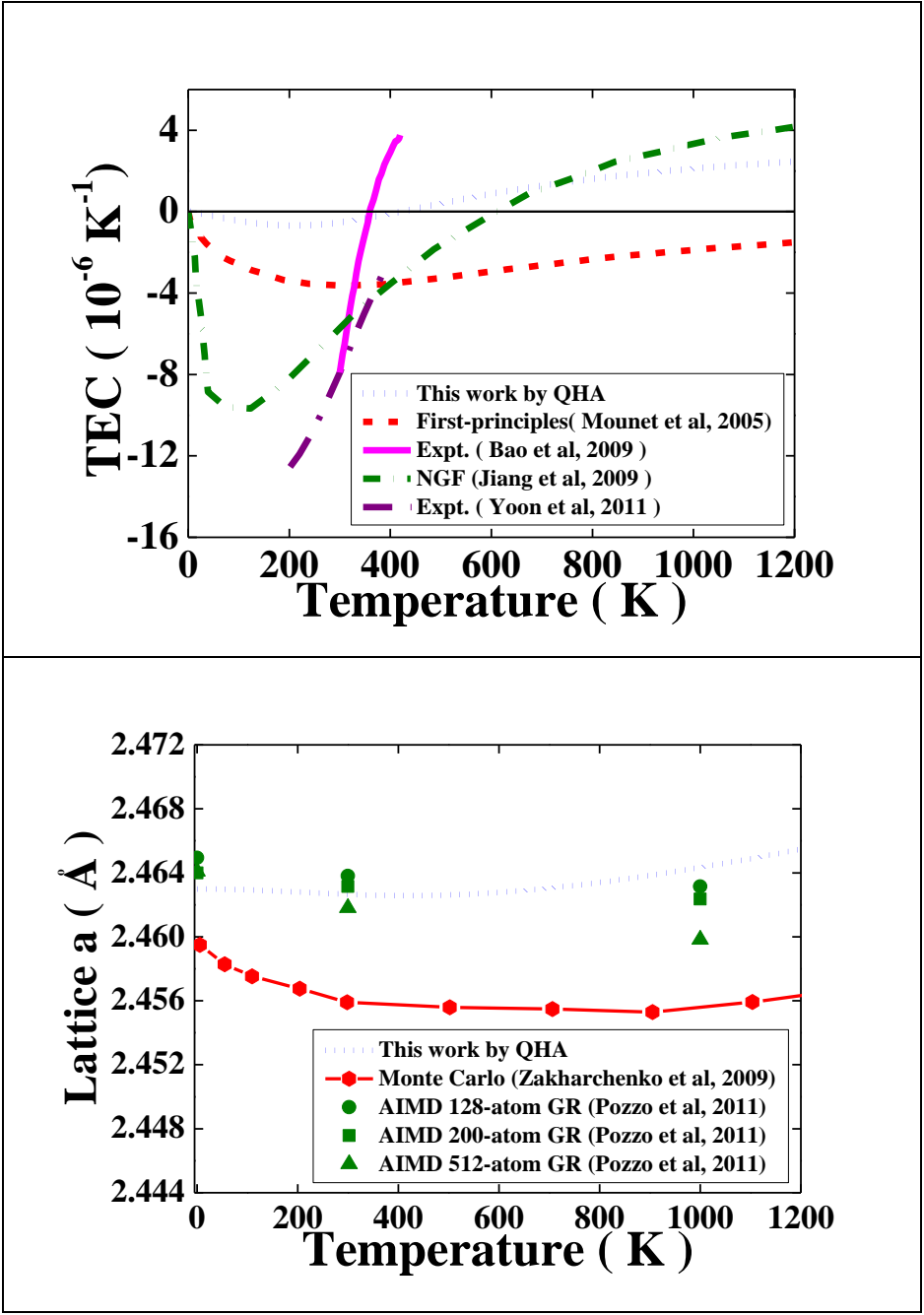

**Figure 2**. (a) Thermal expansion coefficient (TEC) of graphene (GR) calculated in this work under the QHA model (blue dot line); comparisons are also made with TECs calculated by first-principles from Mounet et al's work (red short-dash line) [31]; measured in Bao et al's work (pink solid line) [34], calculated by employing the non-equilibrium Green's function approach from Jiang et al's work [32] (green dash-dot line); measured by Raman Spectroscopy from Yoon et al's work [36] (purple dash-dot line); (b) Comparisons for the relationship of temperature dependent lattice parameter *a* versus temperature in the range from 0 K to 1200 K with those calculated by Monte Carlo approach from Zakharchenko et al's work ( red scatter solid line) [33] and calculated by the AIMD from Pozzo et al' work (green scatters) [35].

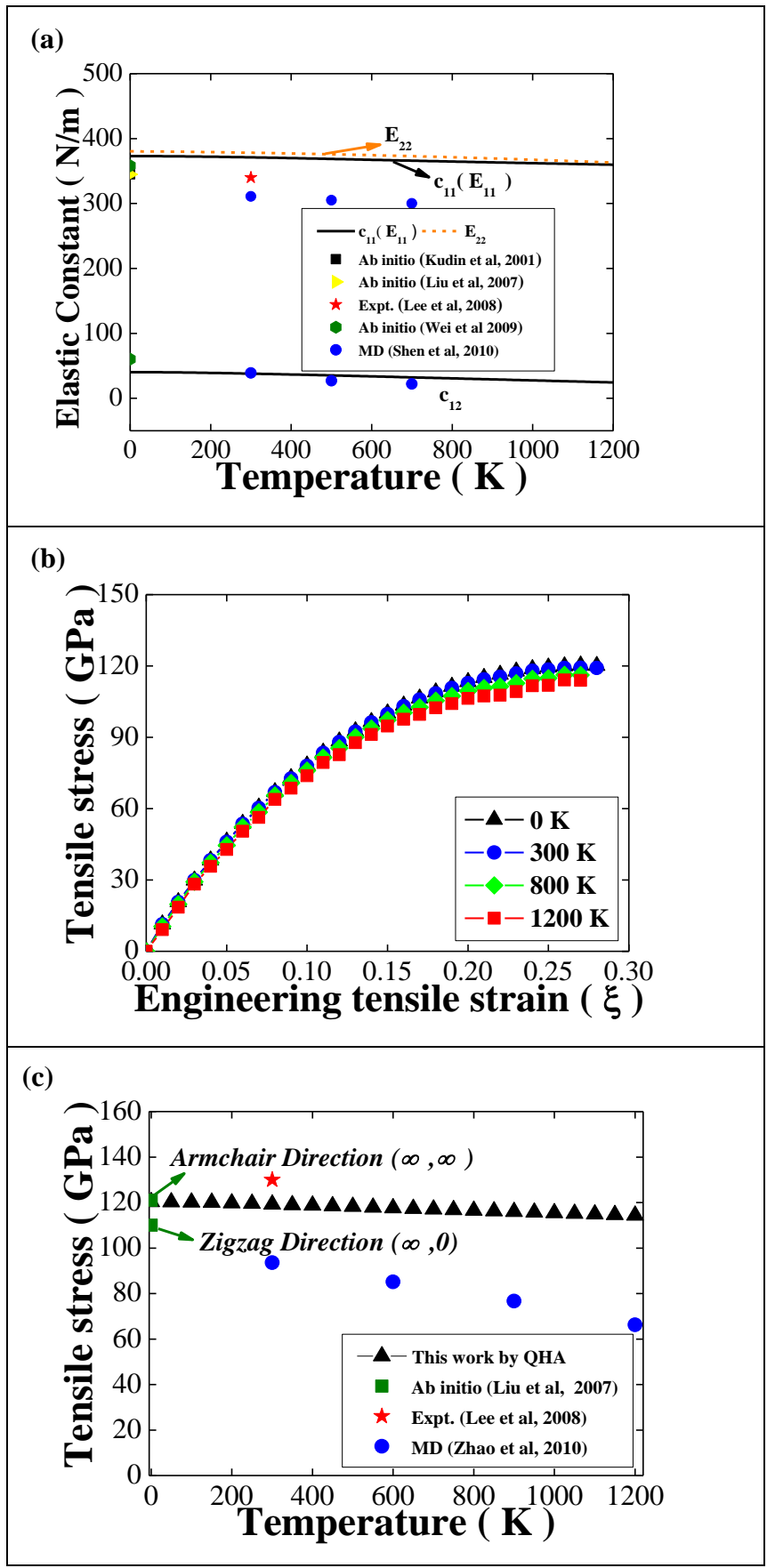

**Figure 3**. (a) Comparison for temperature dependent elastic constant (TDEC), Young modulus $E_{11}$ ($c_{11}$) and $E_{22}$ of graphene versus temperature in the range from 0 K to 1200 K (black solid line) with the Young modulus calculated by the DFT from Kudin et al's work [7] (black square); Young modulus calculated by the DFT from Liu et al's work [8] (yellow triangle); Young modulus measured at room temperature by Lee et al [3] (red star); $c_{11}$, $c_{12}$ at 0 K calculated by the DFT from Wei et al's work [11] (green hexagon); $c_{11}$, $c_{12}$ calculated by the MD from Shen et al's work [10] (blue round); $c_{11}$, $c_{12}$ (black solid line), Young modulus $E_{11}$ ($c_{11}$) and $E_{22}$ (origin short-dash line) in this work. (b) Tensile stress versus equivalent engineering strain for graphene along the [1, 0] direction at 0 K, 300 K, 800 K and 1200 K. (c) Comparisons for the temperature dependent ultimate strength (TDUS) for graphene along the [1, 0] direction at various temperatures in the range from 0 K to 1200 K in this work (black triangle) with those calculated by the DFT from Liu et al's work [8] (yellow triangle), measured at room temperature by Lee et al [3] (red star), calculated by the MD from Zhao et al's work [14] (blue triangle).

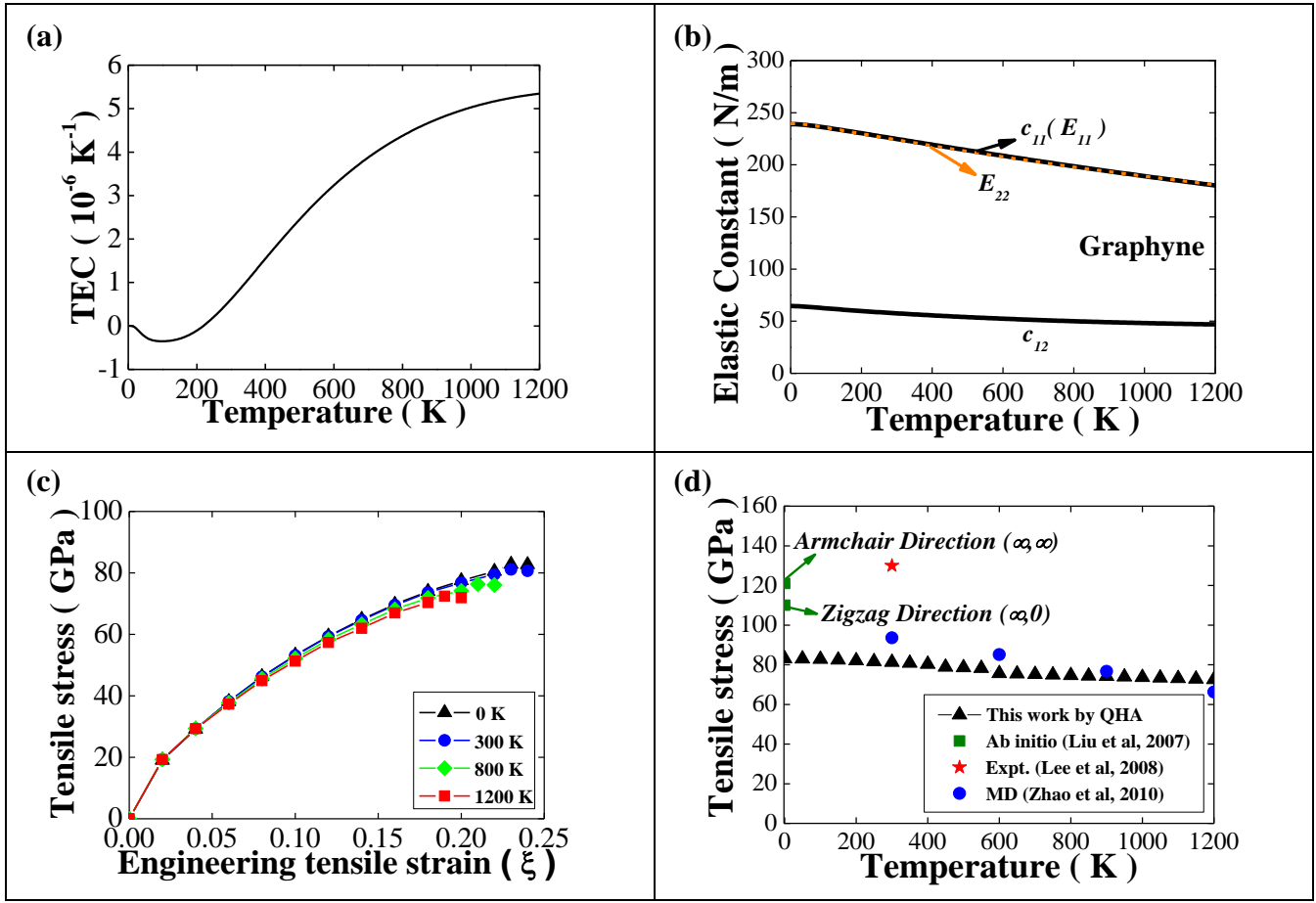

**Figure 4**. (a) In-plane linear thermal expansion coefficient (TEC) of hexagonal graphyne calculated in this work under the QHA model (black solid line). (b) Temperature-dependent elastic constants $c_{11}$, $c_{12}$ (black solid line), Young modulus $E_{11}(c_{11})$ and $E_{22}$ (origin short-dash line) for graphyne versus temperature in the range from 0 K to 1200 K. (c) Tensile stress versus equivalent engineering strain for graphyne along the x-axis direction at 0 K, 300 K, 800 K and 1200 K. (d) Comparisons for temperature-dependent ultimate strength (TDUS) for graphyne along the x-axis at various temperatures varying from 0 K to 1200 K (black triangle) with those calculated by the DFT from Liu et al's work [8] (green triangle), measured at room temperature by Lee et al [3] (red star), calculated by the MD from Zhao et al's work [14] (blue round).

**Table 1.** Mechanical properties of graphene at different temperatures reported from calculations or measurements at different temperatures (K).

| Temp (K) | Method | $c_{11}$ (N/m) | $c_{12}$ (N/m) | Young Modulus | Ultimate Strength (GPa) | Year | References |
|---|---|---|---|---|---|---|---|
| 0 | DFT | | | 345 N/m | | 2001 | [7] |
| 0 | DFT | | | 1050 GPa | 110 ($\infty$,0) <br> 121 ($\infty$,$\infty$) | 2007 | [8] |
| 300 | Expt. | | | 340 N/m <br> 1020 GPa | 130 | 2008 | [3] |
| 0 | Ab-initio | 358.1 | 60.4 | | | 2009 | [11] |
| 300 | MD | 306 | 21 | | | | [13] |
| 500 | MD | 304 | 18 | | | | [13] |
| 700 | MD | 300 | 17 | | | | [13] |
| 300 | MD | | | | 93 | 2010 | [14] |
| 600 | MD | | | | 85 | 2010 | [14] |
| 900 | MD | | | | 76 | 2010 | [14] |
| 1200 | MD | | | | 66 | 2010 | [14] |

**Table 2**. Deformation tensors used to calculate the elastic constants for hexagonal graphene lattice and hexagonal graphyne lattice. A linear combination of elastic constants (LCEC) equals to the second order strain derivatives of the Helmholtz free energy under the corresponding deformation modes (or deformation tensors).

| *Type of crystals lattice* | *Name of deformation modes* | *Deformation tensors* | *LCEC* |
|---|---|---|---|
| Hexagonal symmetry | Zigzag axial deformation | $\begin{pmatrix} \sigma & 0 \\ 0 & 0 \end{pmatrix}$ | $C_{11}$ |
| | Hydrostatic planar deformation | $\begin{pmatrix} \sigma & 0 \\ 0 & \sigma \end{pmatrix}$ | $2(C_{11} + C_{12})$ |